\documentclass[12pt]{article}
\usepackage{amsfonts}
\usepackage{amssymb}
\usepackage{graphics,psboxit,amsmath}

\def\hybrid{\topmargin -20pt    \oddsidemargin 0pt
        \headheight 0pt \headsep 0pt
        \textwidth 6.35in       
        \textheight 9.25in       
        \marginparwidth .875in
        \parskip 5pt plus 1pt   \jot = 1.5ex}

\hybrid

\def\baselinestretch{1.2}

\catcode`\@=11

\def\marginnote#1{}
\newcount\hour
\newcount\minute
\newtoks\amorpm
\hour=\time\divide\hour by60
\minute=\time{\multiply\hour by60 \global\advance\minute by-\hour}
\edef\standardtime{{\ifnum\hour<12 \global\amorpm={am}%
        \else\global\amorpm={pm}\advance\hour by-12 \fi
        \ifnum\hour=0 \hour=12 \fi
        \number\hour:\ifnum\minute<10 0\fi\number\minute\the\amorpm}}
\edef\militarytime{\number\hour:\ifnum\minute<10 0\fi\number\minute}

\def\draftlabel#1{{\@bsphack\if@filesw {\let\thepage\relax
   \xdef\@gtempa{\write\@auxout{\string
      \newlabel{#1}{{\@currentlabel}{\thepage}}}}}\@gtempa
   \if@nobreak \ifvmode\nobreak\fi\fi\fi\@esphack}
        \gdef\@eqnlabel{#1}}
\def\@eqnlabel{}
\def\@vacuum{}
\def\draftmarginnote#1{\marginpar{\raggedright\scriptsize\tt#1}}

\def\draft{\oddsidemargin -.5truein
        \def\@oddfoot{\sl preliminary draft \hfil
        \rm\thepage\hfil\sl\today\quad\militarytime}
        \let\@evenfoot\@oddfoot \overfullrule 3pt
        \let\label=\draftlabel
        \let\marginnote=\draftmarginnote
   \def\@eqnnum{(\theequation)\rlap{\kern\marginparsep\tt\@eqnlabel}%
\global\let\@eqnlabel\@vacuum}  }

\def\preprint{\twocolumn\sloppy\flushbottom\parindent 2em
        \leftmargini 2em\leftmarginv .5em\leftmarginvi .5em
        \oddsidemargin -.5in    \evensidemargin -.5in
        \columnsep .4in \footheight 0pt
        \textwidth 10.in        \topmargin  -.4in
        \headheight 12pt \topskip .4in
        \textheight 6.9in \footskip 0pt
        \def\@oddhead{\thepage\hfil\addtocounter{page}{1}\thepage}
        \let\@evenhead\@oddhead \def\@oddfoot{} \def\@evenfoot{} }

\def\numberbysection{\@addtoreset{equation}{section}
        \def\theequation{\thesection.\arabic{equation}}}

\def\underline#1{\relax\ifmmode\@@underline#1\else
        $\@@underline{\hbox{#1}}$\relax\fi}

\def\titlepage{\@restonecolfalse\if@twocolumn\@restonecoltrue\onecolumn
     \else \newpage \fi \thispagestyle{empty}\c@page\z@
        \def\thefootnote{\fnsymbol{footnote}} }

\def\endtitlepage{\if@restonecol\twocolumn \else \newpage \fi
        \def\thefootnote{\arabic{footnote}}
        \setcounter{footnote}{0}}  

\catcode`@=12
\relax

\def\figcap{\section*{Figure Captions\markboth
        {FIGURECAPTIONS}{FIGURECAPTIONS}}\list
        {Figure \arabic{enumi}:\hfill}{\settowidth\labelwidth{Figure
999:}
        \leftmargin\labelwidth
        \advance\leftmargin\labelsep\usecounter{enumi}}}
 \relax
\def\tablecap{\section*{Table Captions\markboth
        {TABLECAPTIONS}{TABLECAPTIONS}}\list
        {Table \arabic{enumi}:\hfill}{\settowidth\labelwidth{Table
999:}
        \leftmargin\labelwidth
        \advance\leftmargin\labelsep\usecounter{enumi}}}
 \relax
\def\reflist{\section*{References\markboth
        {REFLIST}{REFLIST}}\list
        {[\arabic{enumi}]\hfill}{\settowidth\labelwidth{[999]}
        \leftmargin\labelwidth
        \advance\leftmargin\labelsep\usecounter{enumi}}}
 \relax
\makeatletter
\newcounter{pubctr}
\def\publist{\@ifnextchar[{\@publist}{\@@publist}}
\def\@publist[#1]{\list
        {[\arabic{pubctr}]\hfill}{\settowidth\labelwidth{[999]}
        \leftmargin\labelwidth
        \advance\leftmargin\labelsep
        \@nmbrlisttrue\def\@listctr{pubctr}
        \setcounter{pubctr}{#1}\addtocounter{pubctr}{-1}}}
\def\@@publist{\list
        {[\arabic{pubctr}]\hfill}{\settowidth\labelwidth{[999]}
        \leftmargin\labelwidth
        \advance\leftmargin\labelsep
        \@nmbrlisttrue\def\@listctr{pubctr}}}
 \relax
\makeatother
\newskip\humongous \humongous=0pt plus 1000pt minus 1000pt

\newif\ifdtup

\relax

\def\be{\begin{equation}}
\def\ee{\end{equation}}
\def\ba{\begin{eqnarray}}
\def\ea{\end{eqnarray}}

\def\no{\noindent}

\def\IR{\relax{\rm I\kern-.18em R}}
\def\II{\relax{\rm 1\kern-.35em1}}

\def\IR{\relax{\rm I\kern-.18em R}}
\def\inv{^{\raise.15ex\hbox{${\scriptscriptstyle -}$}\kern-.05em 1}}

\begin{document}

\begin{titlepage}
\begin{center}

\hfill CERN-PH-TH/2006-048\\
\vskip -.1 cm
\hfill IFT-UAM/CSIC-06-14\\
\vskip -.1 cm
\hfill hep--th/0603204\\

\vskip .5in

{\LARGE Quantum corrections to the string Bethe ansatz}
\vskip 0.4in

{\bf Rafael Hern\'andez$^1$}\phantom{x} and\phantom{x}
 {\bf Esperanza L\'opez}$^2$ 
\vskip 0.1in

${}^1\!$
Theory Division, CERN\\
CH-1211 Geneva 23, Switzerland\\
{\footnotesize{\tt rafael.hernandez@cern.ch}}

\vskip .2in

${}^2\!$
Departamento de F\'{\i}sica Te\'orica C-XI
and Instituto de F\'{\i}sica Te\'orica  C-XVI\\
Universidad Aut\'onoma de Madrid,
Cantoblanco, 28049 Madrid, Spain\\
{\footnotesize{\tt esperanza.lopez@uam.es}}

\end{center}

\vskip .4in

\centerline{\bf Abstract}
\vskip .1in
\no
One-loop corrections to the energy of semiclassical rotating strings contain 
both analytic and non-analytic terms in the 't Hooft coupling. Analytic 
contributions agree with the prediction from the string Bethe ansatz based
on the classical $S$-matrix, but in order to include non-analytic contributions 
quantum corrections are required. We find a general expression for the first 
quantum correction to the string Bethe ansatz.

\noindent

\vskip .4in
\noindent

\end{titlepage}
\vfill
\eject

\def\baselinestretch{1.2}


\baselineskip 20pt


In recent years impressive precision tests of the AdS/CFT correspondence 
have been performed. Moreover, the ambitious prospect of solving the large $N$ 
limit of ${\cal N}\!=\!4$ supersymmetric Yang-Mills and correspondingly, 
the sigma model that describes Type IIB string theory on $AdS_5 \times S^5$, 
looks now a step closer. The crucial ingredient for these developments 
has been the appearance of integrable structures on both sides of the 
correspondence. A major development was to reinterpret the planar dilatation 
operator of the 
gauge theory as the hamiltonian of a spin chain \cite{MZ}, which up to one-loop
for the complete theory \cite{Beisert,psu} and several loops in
restricted sectors \cite{dilatation}-\cite{Serban} was shown to be 
integrable. Assuming integrability, a long range Bethe ansatz 
was then conjectured to describe all higher loop effects 
except for non-local processes in the chain \cite{longrange}. 
On the string theory side, integrability arises because the classical string sigma 
model on $AdS_5 \times S^5$ admits a Lax representation \cite{Wadia,Polchinski}. The 
integral equations satisfied by the spectral density for the Lax operator are the string 
analogue of the Bethe equations for the gauge theory dilatation operator
in the thermodynamic limit \cite{Kazakov}. Assuming that integrability
survives after quantization, a discrete Bethe ansatz has been
suggested \cite{qBethe}-\cite{completeqBethe} which should determine
the quantum spectrum of the string sigma model.

The structure of the gauge and string long range Bethe ans\"atze is
remarkably similar. In particular the tower of conserved charges
is given by the same expressions in terms of the spectral parameter on both
cases \cite{longrange}.  
But the $S$-matrices differ by a phase \cite{qBethe},
\be
S_{st}(p_j,p_k)=S_{g}(p_j,p_k) \, e^{i \, \theta(p_j,p_k)} \ ,
\label{smatrix}
\ee
given by
\be
\theta(p_j,p_k)=2 \sum_{r=2}^\infty c_r(\lambda) 
\left(\!{\lambda \over 16 \pi^2}\!\right)^{\!r} \!
\Big(q_r(p_j) q_{r+1}(p_k)-q_{r+1}(p_j)q_r(p_k) \Big) \ ,
\label{phase}
\ee
where $\lambda$ is the 't Hooft coupling constant and $q_r(p)$ are the conserved charges 
of the integrable system. In order to recover the integrable structure of the classical 
string the coefficients in (\ref{phase}) must satisfy $c_r(\lambda)\!\rightarrow\! 1$ as 
$\lambda \!\rightarrow \! \infty$. The phase $\theta(p_j,p_k)$ is central to reproduce 
\cite{qBethe} the well known behaviour for the masses of small strings in weakly curved
$AdS_5 \times S^5$, $m^2 \sim \sqrt{\lambda}$ \cite{ml}.
This provides a very strong and simple test of the quantum string Bethe ansatz.

Gauge and string theory sides of the AdS/CFT correspondence are accessible in opposite regimes 
of the coupling constant $\lambda$. Indeed, quantitative comparison between anomalous 
dimensions of gauge theory operators and string energies has only been possible on configurations
whose dynamics is governed by an effective coupling constant $1/{\cal J}\!\equiv\!\lambda/J^2$. 
This parameter can be kept small even if $\lambda$ is large provided the associated 
configuration carries a large enough quantum number $J$ \cite{P}. Perfect agreement has 
been found in this way between the spectrum of long gauge operators and semiclassical string 
to order $\lambda^2$ in all cases analyzed \cite{match,Kazakov} \cite{curves}, including 
also the first quantum corrections \cite{corrections}-\cite{SZZ}. However, in spite of this impressive 
results, disagreement is known to start at order $\lambda^3$ \cite{Callan,dis,Serban}. This problem 
can be traced back to the dressing phase in \eqref{smatrix}. Understanding how $\theta(p_j,p_k)$ 
changes in going from strong to weak coupling is crucial to uncover how the gauge theory can rearrange 
itself in terms of a string theory. The objective of this note is to study the first quantum 
corrections to the limiting value of the phase \eqref{phase}, given by $c_r(\infty)\!=\!1$.
  
\vspace{2 mm}
  
In the following we will compare the leading quantum correction to the energy of 
semiclassical strings derived from the string Bethe ansatz and from a one-loop world-sheet
calculation. One-loop corrections can be obtained from the spectrum of quadratic 
fluctuations around a given classical solution,
\be
E_1= \sum_{n=-\infty}^\infty e(n) \, ,
\label{equantum}
\ee
where $e(n)$ is the sum over frequencies of bosonic and fermionic fluctuations with 
mode number $n$. Finding the spectrum of fluctuations is a difficult problem for generic 
semiclassical solutions, but it has been done for the simplest example: circular strings 
\cite{Frolov}-\cite{Satoh}. We will consider two cases: circular strings rotating in an $S^3$ 
section of $S^5$ ($SU(2)$ sector) and in an $AdS_3\times S^1$ section of $AdS_5 \times S^5$
($SL(2)$ sector). 

In order to evaluate \eqref{equantum}, it is useful to first expand $e(n)$ in terms of the 
effective coupling $1/{\cal J}$ and then perform the sum. But this approach involves some problems. Since 
we are working in a supersymmetric string theory, the sum \eqref{equantum} is finite as 
$n\!\rightarrow\!\infty$ (see \eqref{smallsum} and \eqref{horrorsum} in the appendix). Expanding 
$e(n)$ for fixed $n$ at large ${\cal J}$ produces however divergences at high mode number 
\cite{SZZ,SZ}, indicating that this is not appropriate for the high energy tail of the spectrum. On 
the contrary, expanding $e(n)$ at fixed $x\!=\!n/{\cal J}$ is regular at large $x$ but contains
divergences at $x\!=\!0$. With this alternative expansion, the highest modes are well described 
but problems are now shifted to the lowest part of the spectrum. The way out is to combine 
both expansions \cite{BT}
\be
e(n)=e_1(n)+e_2(n/{\cal J}) \, ,
\label{expansions}
\ee
where $e_1(n)$ and $e_2(n/{\cal J})$ denote the regular terms for the fixed $n$
and $n/{\cal J}$ expansions, respectively. The $e_1$ terms are defined using a simple 
zeta-function regularization, and $e_2$ by subtracting negative powers of $x$.
The important relation \eqref{expansions} has been proved recently using 
a Cauchy integral representation of the frequency sum \cite{Sakuraint}.

The contributions to the quantum corrected energy associated to $e_1$ and
$e_2$ are rather different. From \eqref{smallsum} and \eqref{horrorsum}, it is clear
that $\sum e_1$ contains only even powers of $1/{\cal J}$. Therefore
it is analytic on the coupling constant $\lambda$, as it is the case
for the classical energy \cite{circular}. Up to exponentially suppressed terms, 
$\sum e_2$ can be evaluated by substituting the sum over modes by an 
integral. The simplest example to analyze is that of a $SU(2)$ circular
string with $k\!=\!2m$, leading to \cite{BT,MT,Sakuraint}
\be
\int_{-\infty}^\infty {\cal J} dx \, e_2(x)={1\over \sqrt{{\cal J}^2+m^2}}\left(
m^2+2 {\cal J}^2 \log{{\cal J}^2 \over {\cal J}^2+m^2}-{{\cal J}^2-m^2 \over 2}
\log{{\cal J}^2-m^2 \over {\cal J}^2+m^2} \right) \, .
\label{sumsu2}
\ee
This expression expands in odd powers of $1/{\cal J}$, giving thus 
raise to non-analytic contributions in $\lambda$.\footnote{Circular
strings rotating on $S^3\!\subset \! S^5$ are unstable due to tachyonic modes at
low momentum. Since \eqref{sumsu2} is an effect of the highest tail of the
spectrum, it should not be affected by that problem.} 
This is a generic pattern. For $SL(2)$ circular strings we have \cite{BT}
\be
\int_{-\infty}^\infty {\cal J} dx \, e_2(x)=-{(k-m)^3 \, m^3 \over 3 {\cal J}^5}
\left( 1 - {3 k^2-8 k m \over 2 {\cal J}^2} + \dots \right) \, ,
\label{sumsl2}
\ee
where $k$ and $m$ are winding numbers that characterize the 2-spin
circular strings.

From the point of view of the Bethe ansatz for quantum strings, there are
two possible sources of contribution to $E_1$: finite size effects and
quantum corrections to the classical integrable structure. Finite size effects 
provide $1/J$ corrections to the leading order result, obtained using
the thermodynamic limit of the Bethe ansatz. Since the classical energy is 
analytic in $\lambda$ for large ${\cal J}$, this will also be the case for the
finite size corrections. It was shown in \cite{SZZ} that the $S$-matrix
\eqref{smatrix}, with coefficients
$c_r(\lambda)\!=\!1$ in the dressing phase, reproduces the
contribution to $E_1$ from $\sum e_1$ at least 
up to order $1/{\cal J}^6$. This simple choice, trivially consistent
with the classical value $c_r(\infty)\!=\!1$,
does not produce however non-analytic terms in $\lambda$.
This implies that in order to include the contribution from $\sum e_2$,
and assuming that integrability is maintained at the quantum level, it is
necessary to introduce quantum corrections to the string Bethe ansatz.
   
The most general Bethe ansatz for a system with $SU(2|2)$ symmetry was
analyzed in \cite{su22}. The dispersion relation turned to be fixed 
by the symmetry algebra. The only freedom left in the integrability structure was contained in 
an undetermined phase in the scattering matrix. The most general form of the dressing phase of 
the $S$-matrix was studied in \cite{Klose} for an integrable system with $GL(n)$ symmetry. It 
was found to be of the form \eqref{phase}, but with two uncorrelated conserved charges 
$q_r(p_j)$ and $q_s(p_k)$. 
In accordance with these results, it was proposed in \cite{BT} that 
agreement with the sum over frequencies \eqref{equantum}, including non-analytic terms, 
could be restored by generalizing the phase \eqref{phase} to
\be
\theta(p_j,p_k)=2 \sum_{r=2}^\infty \sum_{s=r+1}^{\infty}
c_{r,s}(\lambda) \left({\lambda \over 16 \pi^2}
\right)^{\!\!\!\!{r+s-1 \over 2}} \Big(q_r(p_j) q_{s}(p_k)-q_{s}(p_j)q_r(p_k)
\Big) \, ,
\label{newphase}
\ee
with
\be
c_{r,s}=\delta_{r+1,s} + {1 \over \sqrt{\lambda}} a_{r,s} \, .
\ee
  
The result of this note is a conjecture for the previous coefficients,
\be
a_{r,s} = - 8 \, { (r-1)(s-1) \over (r+s-2)(s-r)} \, ,
\label{coeffs} 
\ee
for odd $r\!+\!s$. Parity invariance implies that $a_{r,s}\!=0$ for even $r\!+\!s$ 
\cite{su22}. The value of the coefficient for the lowest correction term, $a_{2,3}$, 
was already derived in \cite{BT}. We are assuming that the modified dressing phase 
\eqref{newphase} is common for all sectors of the correspondence, as it was the case in 
\cite{Staudacher-S,completeqBethe}. In consonance, it is interesting to stress that 
$e_2(x)$, source of the non-analytical/quantum terms, receives contributions from 
bosonic and fermionic string fluctuations in all $AdS_5 \times S^5$ directions 
\cite{Sakuraint}. However, at the same time this fact renders remarkable that 
its effects can be accommodated in a Bethe ansatz formulation, which can be 
consistently truncated inside each sector. In the following we will provide evidence 
in favor of \eqref{coeffs}.
  
The above generalization of the dressing phase represents a quantum correction to 
the scattering matrix in infinite volume. Therefore the modifications that it produces 
on the energy can be analyzed using the thermodynamic limit of the Bethe ansatz. The 
integral Bethe equations that correspond to including a single correction term are
\ba 
2  \eta - \hspace{-4.7mm} \int_{\cal C} dy {\rho (y) \over x-y}&=&2 \pi \eta \, k_i 
+ {x \over x^2-g^2} \left[ 1+ g^2 \int_{\cal C} dy \, {\rho (y) \over y} 
\left( {1+\eta \over y}-{1-\eta \over x} \right) \right. \label{betheint} \\
&& - \; \left. 2 a_{r,s} \, \epsilon \, g^{r+s-1}
\int_{\cal C} dy \, \rho (y) \left({1 \over x^{r-1} y^{s-1}}-
{1 \over x^{s-1} y^{r-1}} \right) \right] \ , \hspace{.3cm} x\in 
{\cal C}_i \ , 
\nonumber 
\ea
where $\eta=1,-1$ for the $SU(2)$ and $SL(2)$ cases, respectively. For simplicity we 
have denoted the coupling constants that govern the thermodynamic scaling and the 
quantum corrections by $g$ and $\epsilon$, respectively,
\be
g={1 \over 4 \pi {\cal J}} \ , \hspace{1cm} 
\epsilon={1 \over \sqrt{\lambda}} \ ,
\ee
and ${\cal C}= \cup_i \, {\cal C}_i$ represents the set of curves in the complex 
$x$-plane where the Bethe roots condense. Circular strings correspond to solutions 
of the Bethe equations where all roots lie on a single connected curve. In this case 
there is just one mode number $k_i \equiv k$, and \eqref{betheint} can be rewritten as 
an algebraic equation for the resolvent,
\ba
G^2 -2 \pi k \, G - \eta \, G^{(1)} \! &=& \! g^2\, 
\big({G^{(1)}}^2 - 2 \pi k \, G^{(2)}\big)
+g^2 \, (1+\eta) \, \big(G^{(1)} Q_2- G^{(2)} Q_1\big) \nonumber \\
&&-\;2 a_{r,s} \, \eta  \epsilon  \, g^{r+s-1} \,
\big( G^{(r)} Q_s-G^{(s)}Q_r \big) \, ,
\label{resolv}
\ea
where
\be
G(x)=\int_{\cal C} dy {\rho\,(y) \over x-y} \, .
\ee
The infinite tower of conserved charges can be directly read from the expansion of 
the resolvent around $x=0$,
\be
G(x)=-\sum_{n=0}^\infty Q_{n+1} x^n \, , \hspace{1cm}  
Q_n= \int_{\cal C} dy \, {\rho(y) \over y^n} \, .
\ee
The Virasoro constraints imply that the total momentum of the sigma model configuration must 
be a multiple of $2 \pi$: $P\!=\!Q_1\!=\!- 2\pi m$, with $m \in \mathbb{Z}$. The second 
conserved charge determines the energy of the circular string: $E\!=\!J\,(1+2 g^2 Q_2)$. 
Finally, in \eqref{resolv} we have introduced the convenient notation
\be
G^{(r)}(x)=-\sum_{n=r}^\infty Q_{n+1}\, x^{n-r} \, .
\ee

The simplest case to analyze is again $SU(2)$ with $k\!=\!2m$, since for it $G^{(2)}$ cancels 
in equation \eqref{resolv} among the first and second terms of the rhs. The modification 
on the energy caused by the additional term in the dressing phase at leading order in 
$\epsilon$ is then easily derived, 
\be
\delta E=
 a_{r,s} \, g^{r+s} \; 
{ Q_{r+1} Q_s-Q_{s+1} Q_r \over \pi  (1+2 g^2 Q_2)} \ .
\label{evariation}
\ee
This expression can be evaluated at any order in $g$ using the expansion 
of the resolvent, which for this case has the simple form
\be
G(x)= 2 \pi m+{\sqrt{1+(4\pi m g)^2}-\sqrt{1+(4 \pi m x)^2} \over 2(x-g^2/x)} \, .
\ee
Substituting in \eqref{evariation} the values of $a_{r,s}$ proposed in \eqref{coeffs}, 
we get the following contribution to the energy 
\be
\delta E=-{m^6 \over 3 \, {\cal J}^5}+{m^8 \over 3 \, {\cal J}^7}-
{49 \, m^{10} \over 120 \,{\cal J}^9}
+{2 \,m^{12} \over 5 \,{\cal J}^{11}}-
{5749 \,m^{14} \over 13440 \,{\cal J}^{13}}
+ \; \dots \, .
\label{desu}
\ee
This series coincides with the expansion of the one-loop string calculation 
\eqref{sumsu2}, at least up to the order that we have checked: $1/{\cal J}^{101}\,$! 
This matching is already a very strong indication, but however it is not concluding. 
There are $(r\!+\!s\!-\!3)/2$ possible terms in \eqref{newphase} contributing to 
the variation of the energy at order $1/{\cal J}^{r+s}$, and only one number to fit, 
the corresponding coefficient in the expansion of \eqref{sumsu2}. Many different choices 
of $a_{r,s}$ could provide the same agreement. 

To perform a more precise test, we consider next the $SL(2)$ case for general 
values of $k$ and $m$. Equation \eqref{evariation} turns into
\be
(1+2 g^2 Q_2) \, \delta Q_2 + 2 \pi k g^2 \, \delta Q_3 =2 a_{r,s} \epsilon
\, g^{r+s-1}  \big( Q_{r+1} Q_s-Q_{s+1} Q_r \big) \, .
\ee
The second term on the lhs makes the explicit expression for $\delta E$ lengthy and 
cumbersome although straightforward to derive recursively in the coupling constant $g$. 
Substituting again the values given in \eqref{coeffs} for $a_{r,s}$, and with the 
help of Mathematica, we obtain
\be
\delta E=-{(k-m)^3 \,m^3 \over 3\, {\cal J}^5}\; \left[1-
{P_2 \over 2 \, {\cal J}^2}+
{P_{4} \over 40 \,{\cal J}^4}
-{P_{6} \over 80 \,{\cal J}^{6}}+
{P_{8} \over 4480 \,{\cal J}^{8}}
+ \; \dots \right] \, ,
\label{desl}
\ee
where $P_n$ are homogeneous polynomials of degree $n$ in the parameters
$m$ and $k$
\ba
P_2 & = & 3 k^2 -8 k m \, , \nonumber  \\
P_4 & = & 75 k^4 - 455 k^3 m + 679 k^2 m^2 - 153 k m^3 + 29m^4 \, ,\\
P_6 & = & 175 k^6 -1755 k^5 m + 5635 k^4 m^2 -6843k^3 m^3 + 2823 k^2 m^4 
- 562 k m^5 + 2 m^6 \, , \nonumber\\
P_8 & = & 11025 k^8 - 159565 k^7 m + 820785 k^6 m^2 - 1923509 k^5 m^3 
+ 2159033 k^4 m^4 \nonumber \\
&& - 1141813 k^3 m^5 + 
303665 k^2 m^6 - 31753 k m^7 + 2557 m^8 \ .\nonumber
\ea
As before, this is in perfect agreement with the contribution
\eqref{sumsl2} to the sum over string fluctuations. We were only able to 
check the matching to order $1/{\cal J}^{13}$. However, contrary to the
previous case, coincidence between the quantum corrected Bethe 
ansatz and the one-loop sigma model calculation completely fixes now the free coefficients in the 
phase \eqref{newphase}. Moreover, the correction to the energy of circular strings at order 
$1/{\cal J}^{r+s}$ is a polynomial with $r\!+\!s\!-\!4$ coefficients while the modified 
dressing phase \eqref{newphase} contains only $(r\!+\!s\!-\!3)/2$ terms contributing to that order. 
Therefore the agreement that we have found is a very non-trivial check both for our conjectured 
form of $a_{r,s}$, and for the proposal \cite{BT,su22} that an extended dressing phase 
as in \eqref{newphase} can take care of the first quantum corrections to the string Bethe ansatz.

The quantum correction in the lowest term of the dressing phase, $r,s\!=\!2,3$, 
reproduces the $1/{\cal J}^5$ contribution in \eqref{sumsu2} and \eqref{sumsl2}.
This lowest term is also present in the strong coupling limit of the dressing phase 
that distinguishes the gauge and string $S$-matrices \eqref{smatrix}. These two facts imply 
that the leading discrepancy between the
gauge and string results for the classical energy must be proportional to the
leading non-analytical contribution to the quantum corrected string energy \cite{BT}
\be
-{16 \over 3 \sqrt{\lambda}} \:
\big(E_{st}-E_g \big)=
-{(k-m)^3 \,m^3 \over 3\, {\cal J}^5} + {\cal O}\!\left(
{1/{\cal J}^7}\right) \ ,
\label{mistic}
\ee
where the proportionality factor determines $c_{2,3}(\lambda)\!=\!1\!-\!16/3\sqrt{\lambda}$. 
Equation \eqref{mistic} indicates a direct relation between the quantum corrections to the 
string Bethe ansatz and the gauge/string discrepancy. The negative correction term opens the 
possibility that $c_{2,3}(\lambda)$ could interpolate smoothly between the strong coupling value, 
$c_{2,3}(\infty)\!=\!1$, and zero at weak coupling, suggesting thus a solution of the 
puzzling three-loop discrepancy \cite{BT}.  

The observation \eqref{mistic} does not extend to higher orders in the coupling 
constant. The mismatch between gauge and string classical energies at order $\lambda^5$ 
does not turn out to be proportional to the $1/{\cal J}^7$ term in the
expansion of the integral 
\eqref{sumsl2}. Coincidence would have suggested that a single coefficient running appropriately 
from strong to weak coupling, perhaps $c_{r,r+1}$, could have been enough to cure the 
discrepancy. The fact that this is not the case is thus consistent with the need to include 
all the terms, with $r$ and $s$ uncorrelated, in order to take into account the complete 
quantum corrections. For $s\!\neq \!r\!+\!1$ we have $c_{r,s}(\infty)\!=c_{r,s}(0)\!=\!0$, 
and $c_{r,s}(\lambda)$ different from zero otherwise. Although this situation is more involved 
than a simple monotonic running implying only the coefficients $c_{r,r+1}$, it is clear from the 
simple form of the $a_{r,s}$ in \eqref{coeffs} that all the terms in the series are related. 
It would be extremely important to have a deeper understanding of the structure behind the 
extended dressing phase in order to clarify the nature of the gauge/string discrepancy. 
  
The structure of the dressing phase should play a central role in the construction of 
a quantum sigma model for the the string on $AdS_5 \times S^5$. Recently a parallel 
approach to the problem has been provided by the study of closely related sigma models 
with a known quantum $S$-matrix \cite{Mann}-\cite{KZ}.  The classical limit of these 
models reproduces the thermodynamic Bethe ansatz for the string in \cite{Kazakov}. 
In addition, the sigma model considered in \cite{Mann} contains non-analytic terms in 
$\lambda$ associated to quantum effects, although it misses finite size corrections 
as it is defined on a plane.\footnote{For an analysis of corrections to
integrable quantum field theories on a cylinder in the AdS/CFT context
see \cite{Ambjorn}.} The study of these models could 
also illuminate the series in the dressing phase.


\vspace{5mm}
\centerline{\bf Acknowledgments}

The work of R.~H. is supported by a CERN fellowship. The work of E.~L. is supported by 
a Ram\'on y Cajal contract of MCYT and in part by the Spanish DGI under contracts FPA2003-02877 
and FPA2003-04597.


\appendix

\renewcommand{\theequation}{\thesection.\arabic{equation}}
\csname @addtoreset\endcsname{equation}{section}

\section{Appendix}
  
For completeness we include in this appendix the expressions of the sum over 
string frequencies of mode number $n$. For the $SU(2)$ case, with $k=2m$, we have
\cite{Frolov,Frolov2,corrections}
\ba
e(n) & = & \sqrt{1+ {(n+\sqrt{n^2-4 m^2})^2 \over 4({\cal J}^2+m^2)}}
+\sqrt{1+ {n^2-2 m^2 \over {\cal J}^2+m^2}}+
2\sqrt{1+ {n^2\over {\cal J}^2+m^2}} \label{smallsum} \\
&& - \: 4 \sqrt{1+ {n^2-m^2 \over {\cal J}^2+m^2}} \, . \nonumber
\ea
The first contribution comes from bosonic fluctuations along the $S^3$ where the classical 
string is rotating. The second and third terms correspond to bosonic fluctuations on the 
remaining $S^5$ and $AdS_5$ directions, respectively. And the last term comes from the 
fermionic fluctuations. 

For the general $SL(2)$ case, we have the following more involved expressions \cite{Park}
\ba
e(n) & = & {1\over 4\kappa} (\omega_1+\omega_2-\omega_3
-\omega_4)+{1\over \kappa}\sqrt{n^2+\kappa^2}+
{2\over \kappa}\sqrt{n^2+{\cal J}^2-m^2} \label{horrorsum} \\
&& -{2\over \kappa}\sqrt{(n-\gamma)^2+{1 \over 2}(\kappa^2+{\cal J}^2-m^2)}
-{2\over \kappa}\sqrt{(n+\gamma)^2+{1 \over 2}(\kappa^2+{\cal J}^2-m^2)} 
\nonumber \ .
\ea
The first term describes the contribution from the bosonic fluctuations along the $AdS_3$ 
section where the circular string is rotating. The second and third terms come from the 
bosonic fluctuations along the remaining $AdS_5$ and $S^5$ directions, respectively. And the 
last two terms correspond to the fermionic fluctuations. The frequencies $\omega_i$
are solutions of the quartic equation
\be
(\omega^2-n^2)^2+4 r^2 \kappa^2 \omega^2 -4(1+r^2)
(\omega \sqrt{\kappa^2+k^2}-k n)^2=0 \, ,
\ee
ordered in decreasing magnitude. The remaining quantities that appear in 
\eqref{horrorsum} are defined through
\ba
r^2 &=& -{{\cal J}m \over k \sqrt{\kappa^2+k^2}} \, ,\\
\gamma &=& {\kappa m \over \sqrt{\kappa^2+k^2}} 
{\kappa^2-{\cal J}^2+k^2 \over \kappa^2-{\cal J}^2+m^2} \nonumber \, .
\ea
Finally, the parameter $\kappa$ can be determined from 
\be
(\kappa^2-{\cal J}^2-m^2) \sqrt{\kappa^2+k^2} +2 {\cal J}k m=0 \, .
\ee



\end{document}